\documentclass[twocolumn,secnumarabic,amssymb, nobibnotes, aps, prx, superscriptaddress]{revtex4-1}

\usepackage{soul}
\usepackage{graphicx}
\usepackage{graphics}
\usepackage{xcolor}
\usepackage[normalem]{ulem}
\usepackage{amsmath}
\usepackage{scalerel}
\usepackage{float}

\def\={\,=\,}

\def\msquare{\mathord{\scalerel*{\Box}{gX}}}

\begin{document}

%\Initializebibliographystyle{aipauth4-1}

\title{
Zero-bias crossings and peculiar Shapiro maps in graphene Josephson junctions}

\author{T. F. Q. Larson} 
\affiliation{Department of Physics, Duke University, Durham, NC 27708, USA.}
\author{L. Zhao}   
\affiliation{Department of Physics, Duke University, Durham, NC 27708, USA.}
\author{E. G. Arnault}   
\affiliation{Department of Physics, Duke University, Durham, NC 27708, USA.}
\author{M. T. Wei}   
\affiliation{Department of Physics, Duke University, Durham, NC 27708, USA.}
\author{A. Seredinski}   
\affiliation{Department of Physics, Duke University, Durham, NC 27708, USA.}
\author{H. Li}   
\affiliation{Department of Physics and Astronomy, Appalachian State University, Boone, NC 28607, USA.}
\author{K. Watanabe}
\author{T. Taniguchi}
\affiliation{Advanced Materials Laboratory, National Institute for Materials Science, 1-1 Namiki, Tsukuba, 305-0044, Japan.}
\author{F. Amet} 
\affiliation{Department of Physics and Astronomy, Appalachian State University, Boone, NC 28607, USA.}
\author{G. Finkelstein}
\affiliation{Department of Physics, Duke University, Durham, NC 27708, USA.}

\date{\today}%

\begin{abstract} 
%Josephson junctions subject to external periodic drive manifest the inverse AC Josephson effect: the phase across a junction locks to the external frequency. The resulting steady ramping of the phase produces a fixed voltage across the junction, $V=n\hbar \omega /2e$, where $n$ counts the number of $2\pi$ periods by which the phase progressed over one period of excitation. The phase locking persists over a range of applied DC current which results in the I-V curve exhibiting quantized voltage plateaus known as ``Shapiro steps''. Here, we investigate the Shapiro steps in graphene-based junction. A wide variety of patterns could be obtained, depending on the carrier density, temperature, RF frequency, and magnetic field. 
%\st{Inverse AC Josephson effect manifests itself in the formation of the ``Shapiro steps'' of quantized voltage in Josephson junctions subject to RF radiation.} 
The AC Josephson effect manifests itself in the form of ``Shapiro steps'' of quantized voltage in Josephson junctions subject to RF radiation.  
%OR: ``Shapiro steps'' of quantized voltage are observed in Josephson junctions subject to RF radiation.  
This effect presents an early example of a driven-dissipative quantum phenomenon and is presently utilized in primary voltage standards. Shapiro steps have also become one of the standard tools to probe junctions made in a variety of novel materials. Here, we study Shapiro steps in a widely tunable graphene-based Josephson junction. We investigate the variety of patterns that can be obtained in this well-understood system depending on the carrier density, temperature, RF frequency, and magnetic field. Although the patterns of Shapiro steps can change drastically when just one parameter is varied, the overall trends can be understood and the behaviors straightforwardly simulated. The resulting understanding may help in interpreting similar measurements in more complex materials.  

\end{abstract}

\maketitle

\newpage

Josephson junctions subject to an external RF radiation demonstrate the inverse AC Josephson effect: the phase difference across the junction locks to the external frequency ~\cite{JosephsonRMP1964}. As a result, the phase steadily ramps with time, and the $I-V$ curves form ``Shapiro steps'' of quantized voltage $V=n\hbar \omega /2e$, where $n$ counts the number of $2\pi$ periods by which the phase progressed over one period of excitation ~\cite{Shapiro1963}. The exact mechanisms of the phase locking and its stability were investigated in detail in the 1980s~\cite{KautzRev1996}. The extremely precise voltage quantization of the steps is presently utilized in primary voltage standards ~\cite{hamiltonJosephsonVoltageStandards2000}.

Recently, interest in topological Josephson junctions have reinvigorated the use of the AC Josephson effect as a tool to probe a junction's current-phase relation (CPR)~\cite{Kwon2004,FuKane2009}. Missing steps and residual supercurrent associated with the anomalous CPR are some of the signatures which have been explored~\cite{Rokhinson2012,Wiedenmann2016,Bocquillon2017,Li4pi2018,Calvez2019}. Many of these studies are performed at relatively low power and frequency; in this regime the measured maps are significantly different from the textbook ``Bessel function'' patterns even in topologically trivial junctions~\cite{KautzRev1996}.
%\AS{Maybe: Early studies of Shapiro steps focused on Josephson junctions with Al leads, which host a relatively small superconducting gap of X. Much of the recent work in topological systems uses leads with larger gaps on the order of Y, such as Nb, NbTiN, and MoRe.}
%\TL{The pattern of Shapiro steps shown by these junctions sometimes lie outside of the Bessel function regime due simply to junction parameters.}
%While junctions often use Al leads, which have a relatively small gap, a fair amount of junctions use larger gap leads (such as Nb, NbTiN and MoRe)\TL{Still need citations/ to figure out what I'm doing with this sentence}. %owing to improved availability, a range of work functions and other material advantages.[citations]  
%\st{The }Large gaps often result in a pattern of Shapiro steps that lie outside of the textbook Bessel function regime~\cite{Tinkham}. 
In this paper, we study the inverse Josephson effect in graphene-based superconductor-normal-superconductor (SNS) junctions. Shapiro steps in this topologically trivial material have been previously explored~\cite{Heersche2007,Komatsu2012,LeeVertJunction2015,kalantreAnomalousPhaseDynamics2019} and used as a reference in the study of topological junctions~\cite{WiedenmannSupp}. We show that a variety of patterns can be obtained within the same junction by tuning the gate voltage and magnetic field. Both the Bessel function regime and the strongly hysteretic regime with ``zero bias crossings'' are observed. We directly simulate the observed patterns using an extension of the resistively and capacitively shunted junction (RCSJ) model and qualitatively explain the observed trends. 

% [Review]. The theory describes the formation of the Floquet bands, which are similar to the Bloch bands in periodic lattices and can even demonstrate topologically non-trivial behavior~\cite{Harper}. \st{This interest to this subject is primarily motivated by the recent advances in the field of optically pumped quantum gases [some review].} Interestingly, one well-known example of an AC driven periodic quantum system dates back to the 60's:

%, in which the system at zero applied DC current spontaneously breaks symmetry by developing a quantized voltage of either $+ \hbar \omega /2e$ or $- \hbar \omega /2e$. The patterns are reproduced by straightforward modeling and can be qualitatively understood in terms of the few dimensional parameters that govern the system dynamics.

\begin{figure*} [t]
    \centering
    \includegraphics[width=\textwidth]{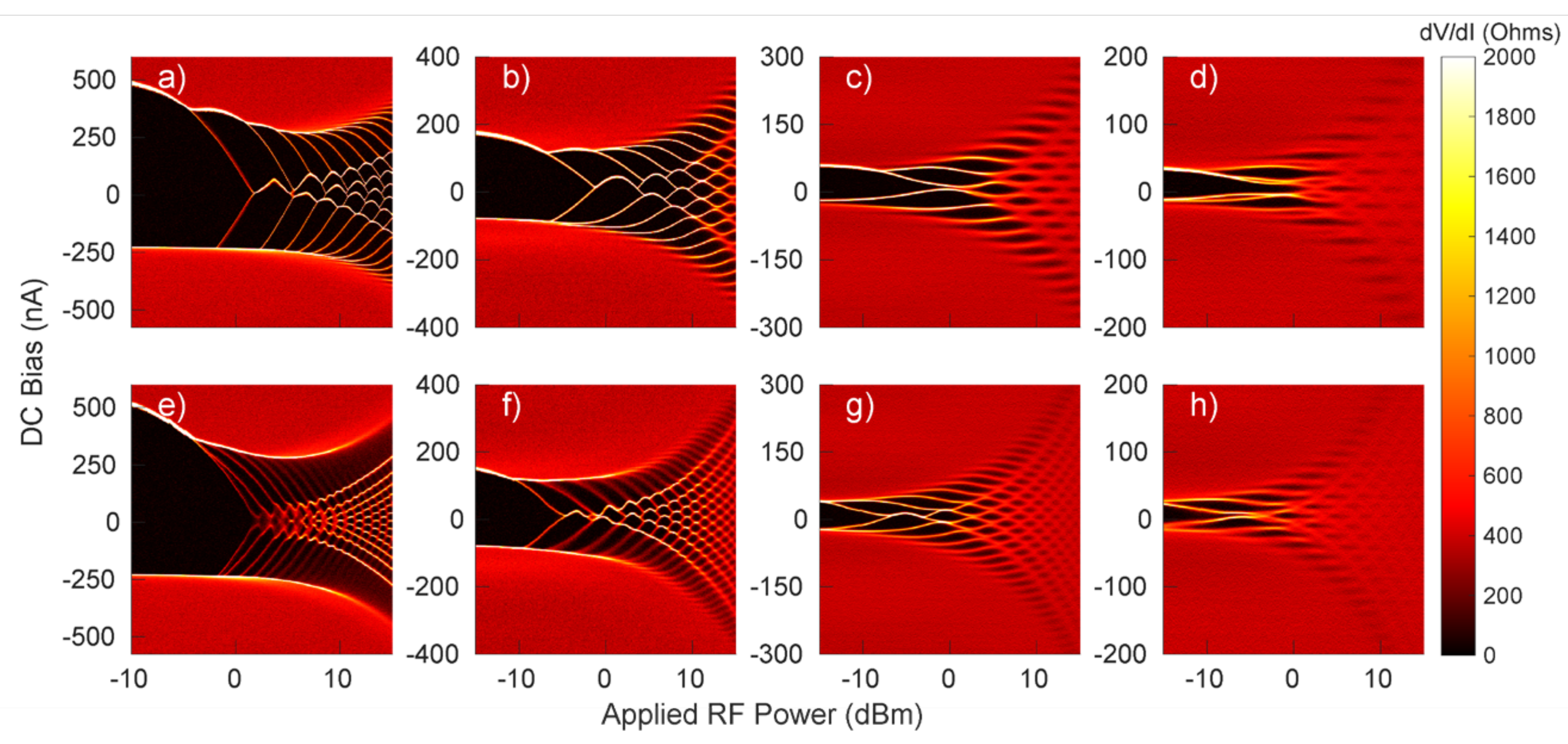}
    \caption{The maps of differential resistance showing Shapiro steps as a function of the DC bias current, $I$, and RF power, $P_{RF}$. The gate voltage for this figure through Figure 4 is set at $V_G= +0.45$ V as measured from the Dirac peak ($V_D= -10$ mV is negligible in this sample). The maps are measured at two frequencies (top row: 5 GHz, bottom row: 3 GHz) and different critical currents (left to right: $I_{S}$ = 650, 240, 80 and 35 nA), as tuned by perpendicular magnetic field). An important dimensional parameter controlling the overall behavior is $\Omega=\omega/\omega_p \propto \omega / \sqrt{I_C}$. 
The pairs of panels (a and f), (b and g), (c and h) correspond to roughy equal values of $\Omega$ and therefore appear similar. We observe the expected trends, according to which at high $\Omega$ (right panels) the plateaus are centered at fixed voltages and their vertical extent is described by the Bessel functions. In the opposite limit of low $\Omega$ (left panels), many features of the maps become hysteretic, and the $n \neq 0$ plateaus cross zero bias. 
%and several unusual features such as a missing first step become apparent.  %Surprisingly, the unquantized voltage and large noise associated with chaotic dynamics in Josephson junctions is not observed in this parameter space.
}
    \label{fig1}
\end{figure*}

Our sample is made of exfoliated graphene encapsulated in hexagonal boron nitride~\cite{Dean2010}. The superconducting leads are made by sputtering molybdenum-rhenium alloy~\cite{CaladoMoRe}, which has a relatively large gap of 1.3 meV. The sample is measured in a dilution refrigerator with the sample holder temperature of approximately 100 mK, which depends weakly on the applied RF signal.  The actual sample temperature under the RF drive could be higher~\cite{DeCecco2016}. The GHz drive is coupled by an antenna placed approximately 1 mm away from the sample. The exact value of the RF power reaching the sample is difficult to quantify because of the frequency-dependent coupling between the antenna and the sample, which are not impedance matched. We therefore only list the nominal RF power emitted by the generator at room temperature, which is a common practice in similar experiments. To measure the DC voltage across the sample, we perform multiple DC current bias sweeps while keeping the rest of the parameters fixed. The resulting $I-V$ curves are then averaged and numerically differentiated to obtain the differential resistance $R=dV/dI$.   %\TL{We note that this is similar to techniques used to measure Cooper pair transistors.} 
%We prefer this method to using a lock-in amplifier for measuring $dV/dI$, especially when multiple hysteretic transitions are present. In such cases, our method effectively produces a transition histogram -- the width of $dV/dI$ peak directly represents the width of the transition region between two voltage plateaus.

%\TL{[[In our defense its also hard because of the antenna and unmatched sample.  Do we need to justify this or can it be assumed?]]}. 
%\TL{ \st{We study in detail a single ballistic graphene Josephson junction.}} 

We start by comparing patterns of Shapiro steps at frequencies of 3 and 5 GHz and several values of magnetic field applied perpendicular to the junction (Figure 1). Magnetic field allows us to conveniently vary the critical current $I_C$ without changing other junction parameters. The maps in Figure 1 present the differential resistance $R$; the dark regions correspond to the steps of constant voltage, and the narrow bright lines correspond to the transitions between these steps. The current bias is swept from negative to positive, resulting in pronounced hysteresis in some of the features. For example, at low applied power, the central dark region corresponding to supercurrent appears asymmetric, with the switching current at positive bias being significantly larger than the retrapping current at negative bias. Many of the step transitions show hysteresis as well.

Following Ref.~\cite{KautzRev1996}, we introduce a convenient dimensionless parameter, $\Omega=\omega/\omega_p$, which is crucial in determining the pattern of Shapiro steps. (Here, $\omega$ is the RF drive frequency and $\omega_P=\sqrt{2 e I_C/\hbar C}$ is the plasma frequency of the junction.) $\Omega$ grows left to right and bottom to top in Figure 1. Shapiro patterns measured at different $\omega$ and $I_C$ but comparable $\Omega$, (see the three pairs (a and f), (b and g), and (c and h) in Figure 1) demonstrate qualitative similarity. 
%\TL{despite having slightly different quality factors $Q$}. 

For the smallest $I_C$ (highest $\Omega$, Figures 1c, d, and h), the pattern of Shapiro steps follows the Bessel function dependence~\cite{Tinkham}. In this regime, the extent of the steps in the bias direction is roughly proportional to the Bessel function $J_n(i)$, where $i=I_{RF}/I_C$ is the ratio of the applied RF current to the critical current~\cite{Russer1972}. The steps are centered at $I= V_n /R_j$, where $R_j$ is the effective DC shunt resistance of the junction. Experimentally, we can extract the effective value of the shunt resistance, $R_j \approx$ 250 Ohms, independent of $I_C$ through Figure 1. Note that this value is comparable, but slightly smaller than the normal resistance of the junction $R_{N} \approx$ 450 Ohms~\cite{Tinkham}. 

 %It is known that the Bessel function regime is achieved in several regions of the phase space, including both $\Omega \gg 1$ regime and the regime of high RF current, $i \gg 1$.~\cite{Kautz} Indeed, it is visible that the centers of the plateaus are aligned for $\Omega \gg 1$ (Figures 1c, d and 1h) and for $i \gg 1$ (all panels). 
%the location of Shapiro steps also approaches $V_n /R_0$ in the middle panels of Figure 1 when the RF power is sufficiently high. Importantly,
%As the critical current increases on the left panels of Figure 1, the patterns change \TL{due to the coexistance of multiple stable steps for given bias values}, and the plateaus are no longer centered around $V_n /R_0$. Instead, starting at small applied power, the $n=\pm 1$ steps split from the switching and retrapping transitions, followed by the steps with with higher $n$The steps descend diagonally toward zero as the RF power is increased; the $n=\pm 1$ steps cross zero bias, and an intricate net of transitions develops at higher RF powers (Figures 1a, b, f and g).. 

As the critical current increases on the left panels of Figure 1, the patterns change due to the coexistence of multiple stable steps for a given bias value~\cite{kalantreAnomalousPhaseDynamics2019}.  The plateaus are no longer centered around a fixed current bias of $V_n /R_j$, but instead emerge sequentially from the normal state boundary and diagonally descend toward zero bias. At high RF power multiple step boundaries intersect, resulting in an intricate net of transitions (Figures 1a, b, f and g). Finally, for the lowest $\Omega$ (Figure 1e), the $\pm 1$ steps no longer reach zero upon the first approach, and the $I-V$ curves show a pronounced region of non-quantized voltage close to zero bias ($P_{RF}$ between 0 and 3 dBm). 

We now concentrate on the parameters of Figure 1a which is reproduced in Figure 2a. Figure 2c shows a line cut extracted from Figure 2a (blue), as well as a similar line measured for the opposite sweep direction (red). 
%an equivalent line cut extracted from the map where the DC bias is swept in the opposite direction.  
This confirms that the asymmetric features seen in Figure 1 are indeed due to hysteresis, and that for many parameter values multiple solutions are simultaneously stable. Figure 2b is taken under the same conditions, but at a higher temperature ($T$=1.5 K). At this temperature, the hysteresis of the Shapiro features is nearly gone, %presumably suppressed by fluctuations overcoming the inertia of the system, 
and a regular pattern emerges, resembling a distorted honeycomb. 
%\st{Two regimes are clearly visible, separated by the dashed  lines: (I) the higher DC bias regions, in which the Shapiro steps first split from $I_S$ and $I_R$, and (II) the roughly triangular region at lower DC bias, in which the transitions form a regular honeycomb net.} As we discuss below, in the former regime, the phase steadily evolves across $n$ minima of the ``washboard potential'' per cycle of RF excitation~\cite{Tinkham}. In the latter regime, the phase evolution is not monotonic, and while its total change is still equal to $2\pi n$ per cycle, the phase retraces some of the potential minima back and forth. 

\begin{figure}[t]
    \centering
    \includegraphics[scale = 0.3]{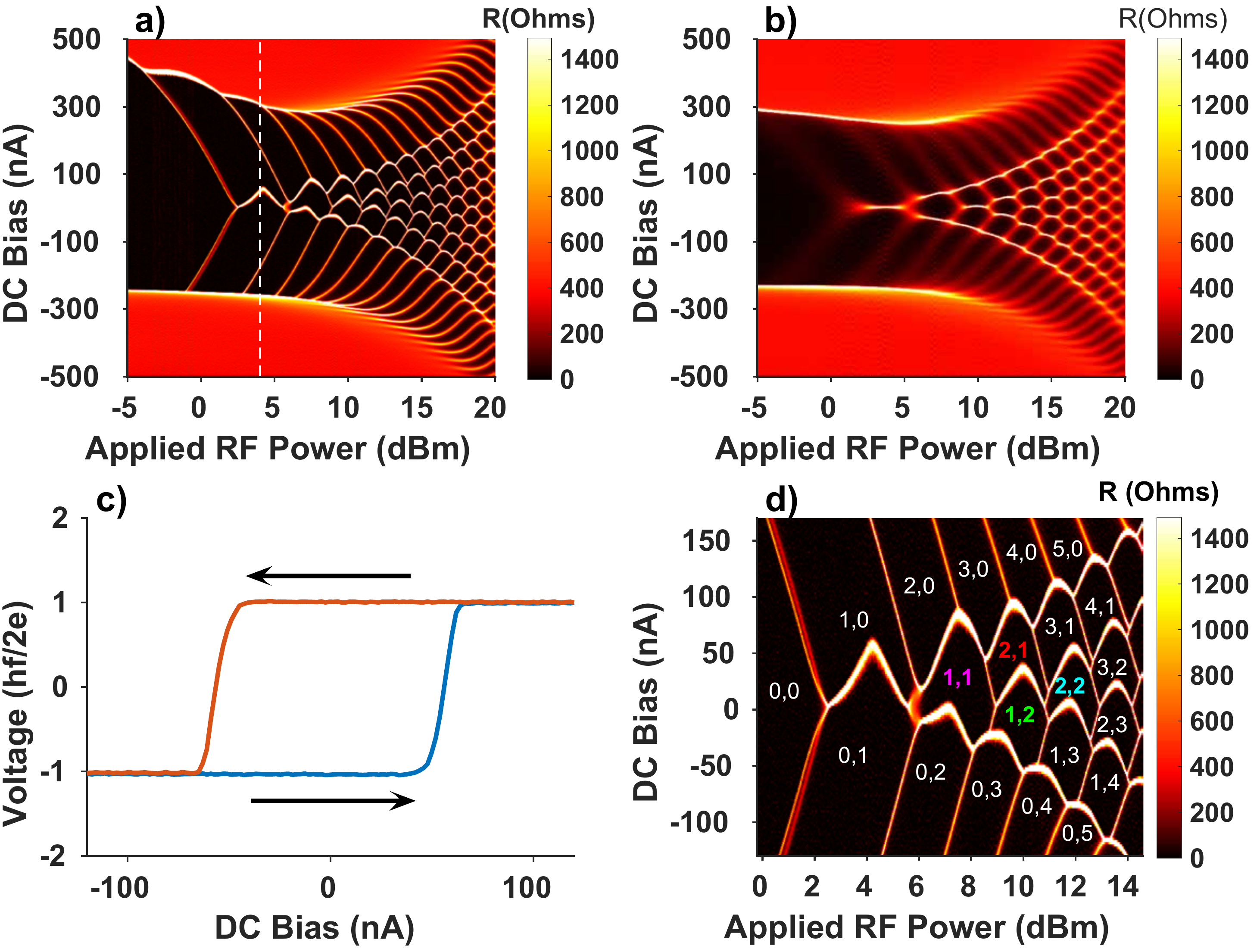}
    \caption{a) Differential resistance as a function of $I$ and $P_{RF}$ reproducing Figure 1a ($f= 5$ GHz, $I_S=$ 650 nA, sample holder temperature $T=$ 100 mK). b) A map identical to (a) but measured at $T=1.5$ K, at which point the hysteresis is largely suppressed. %Regimes (I) and (II) are described in the text. 
    c) A cut through the map (a) taken at $P_{RF}=4$ dBm (at the dashed line), which shows the hysteretic switching between the $n=1$ and $n=-1$ steps depending on the sweep direction. d) A zoom of map (a), with the different plateaus labeled by $(p,q)$ as described in the text.}
    %To further emphasize the hysteresis, (d) shows a symmetrized version of map (a), in which the differential resistances measured for both sweep directions are added together. While the features corresponding to regime (II) of map (b) are hysteretic, the plateau boundaries in the regime (I) cleanly overlap. }
    \label{fig2}
\end{figure}

\begin{figure}[ht]
\centering
\includegraphics[scale=0.3]{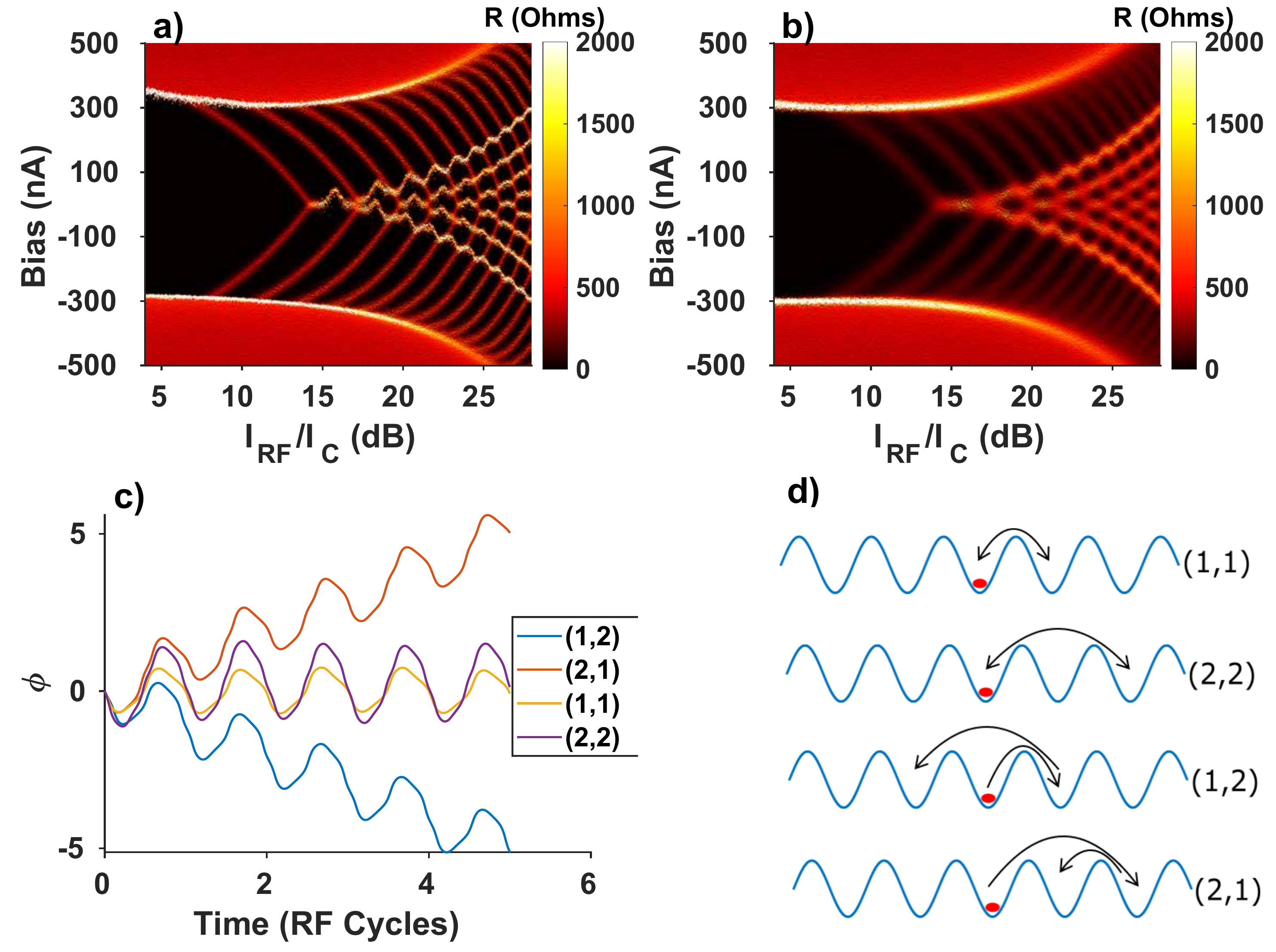}
\caption{a,b) Simulation of $dV/dI$ for different noise levels, to be compared to $R_L=50$ Ohms, $R_j=300$ Ohms, $I_c=600$ nA, $C_{0} = 2.5$ pF and $f=5$ GHz.  To reproduce the experiment, the simulation starts at the lowest DC bias, averages over 500 RF cycles, and then uses the final values of $\phi$ and $\frac{d\phi}{dt}$ as the initial conditions for the next value of bias. Ten bias sweeps are produced in this manner and then averaged to reduce noise. c) Numerical traces of $\phi(t)$ on various plateaus labeled by the pairs of $(p,q)$ (see text). d) Schematic of the washboard potential and the four types of phase evolution corresponding to (c). The top two schematics represent different forms of $n=0$, while the bottom two both show $n=\pm1$.} 
\label{sim2}
\end{figure}

%Returning to Figure 2a, we now recognize a similar pattern, which is distorted by hysteresis: the current is swept negative to positive, resulting in upward ``arching'' of the transition lines. 
%\st{To highlight the hysteresis, Figure 2d shows the symmetrized map where the differential resistance map of Figure 2a is averaged with the measurement for the opposite sweep direction.}  
%The hysteresis can be further highlighted by summing the differential resistance measured for opposite sweep directions (see supplemental Figure S1). We find that the transitions in region II are particularly susceptible to hysteresis; the transitions in region I are not hysteretic at 5 GHz but become hysteretic as well at higher frequencies (Supplementary Figure S1). 

Figures 3a,b show numerical simulations which reproduce most of the features in Figures 2a,b. The simulations use the current-biased model of Ref.~\cite{Russer1972} appropriately modified to account for the capacitance and resistance of the connected pads and leads. Similar to the measurement, time evolution starts at negative bias. We then numerically evolve the equation and average the voltage once the initial transients settle. The final values of $\phi$ and $\frac{d\phi}{dt}$ are then used as initial conditions for the next bias point.  Finally, the sweeps are repeated multiple times and averaged, as is done in the experiment (supplementary material).

These simulations allow us to trace the time evolution of the phase within each cell of the Shapiro map. The examples of the $\phi(t)$ are shown in Figure 3c for several neighboring cells. By analyzing these traces, a rather simple qualitative picture emerges, represented schematically in Figure 3d: For each cycle of RF excitation, the phase progresses over $p$ minima of the washboard potential and then retraces $q$ of them backward. The overall change of phase is $2\pi (p-q)$, and the index of the resulting Shapiro step is $n=p-q$. This behavior has been previously identified in the Bessel function regime~\cite{KautzRev1996,ThorneBardeenpt2}. In Figure 2d, we zoom in on the data of Figure 2a and label select cells by their $(p,q)$ indexes. Note that in the resulting regular pattern, each cell in the central part of the map has 6 neighbors. The two neighbors in the vertical direction have the same total number $p+q$ while n differs by 2. The four neighbors on the left/right have either $p$ or $q$ decreased/increased by 1.  

\begin{figure}[b!]
    \includegraphics[scale = 0.3]{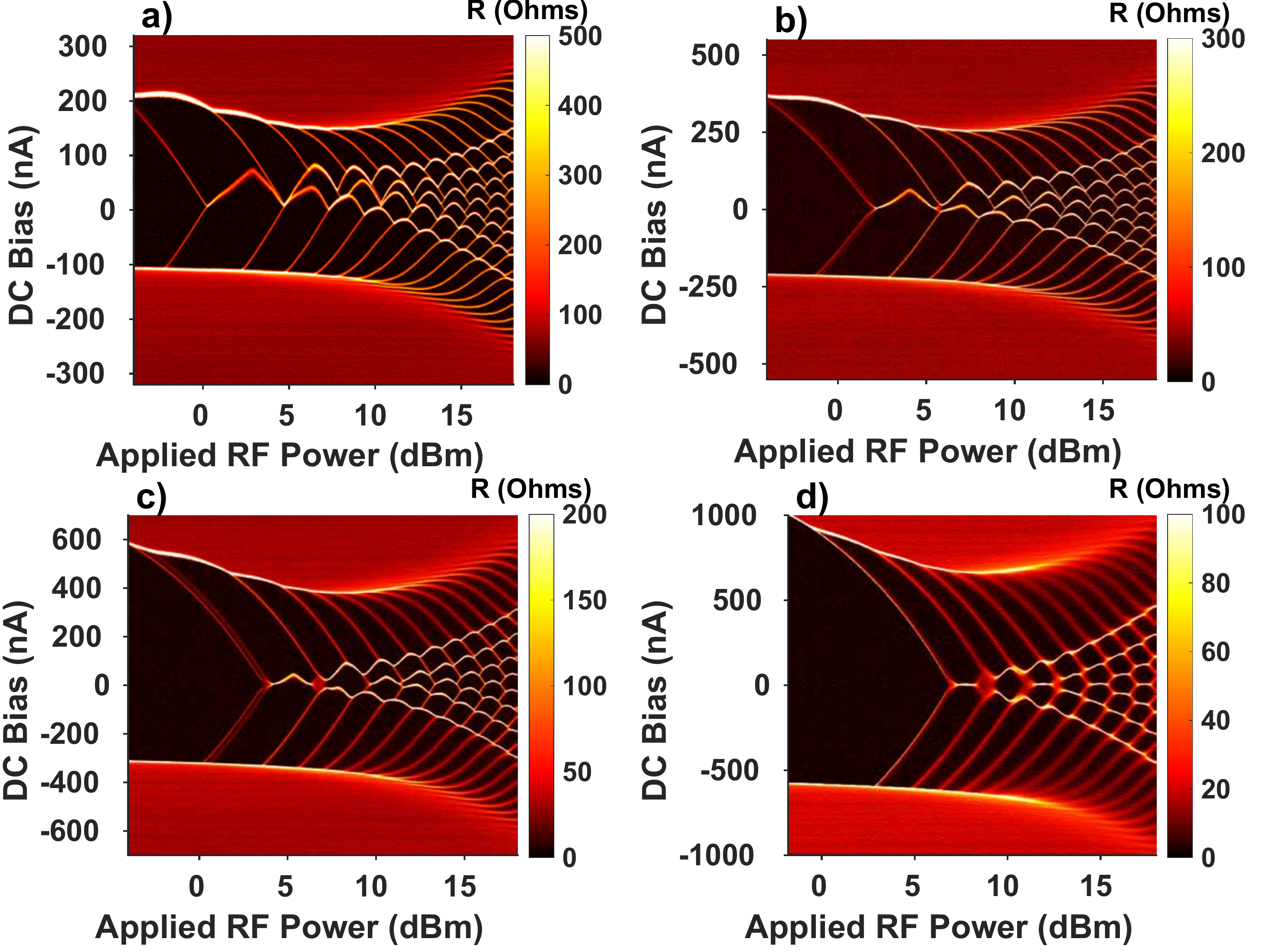}
\caption{a-d) Shapiro maps for gate voltages of 0, 0.375, 0.75 and 1.125 V, respectively. Near the Dirac peak, the hysteresis is the most pronounced, with a gradual decline in hysteresis at higher gate voltages. The higher doping both lowers the junction resistance $R_j$ (resulting in higher dissipation) and increases the critical current.}
\end{figure}

%We note that this is also true in the Bessel function regime, so that for the $m$-th lobe of the $n$-th step the phase progresses by $n+m$ periods and then retraces $m$ periods. 
% up to behavior in the transition between steps, although this point seems to be unnoted in the literature.
%For the $m$-th cell of the $n$-th Shapiro step, labeled as $(n,m)$ 

Finally, we look at the gate voltage dependence of the maps in Figure 4.  While gate voltage influences many parameters,
%including the $I_C$, $R_N$, and their product~\cite{Ivan}, it is assumed that 
the most significant effect is on $I_C$ and $R_{j}$, the latter decreasing by about a factor of roughly 4 between Figures 4a and 4d. We find that near the Dirac peak the hysteresis is very large (Figure 4a), 
%\st{which we attribute to the larger value of $R_{j}$ resulting in reduced dissipation.} 
while further away from the Dirac peak the hysteresis of the transitions between the plateaus is significantly suppressed (Figure 4d). This suppression is partially explained by increased damping, rather than the purely thermal smearing as observed in Figure 2b. However, in order to reproduce the data measured at larger gate voltage, we find that it is also necessary to increase the simulated noise level. We straightforwardly increase the simulated current noise level $\propto \frac{1}{\sqrt{R_{j}}}$, which describes the thermal noise of $R_{j}$. The resulting maps are reasonably consistent with the experiment (supplemental Figure S3). Noise processes in ballistic SNS junctions require further study in both the equilibrium case and under RF drive.  
%[[Last sentence optional -- do we want somehting like this here or in the supplemental?]]  }\st{This observation may indicate that the noise is not simply determined by the thermal noise of the resistors. The situation is different from the simulations in Figure S2, in which the maps of Figure 1 are reproduced keeping the same level of simulated noise. Since the high gate voltage corresponds to a higher critical current, which should be less sensitive to possible spurious noise sources, we surmise that the additional noise originates intrinsically due to the driven-dissipated nature of the system. is instead explained by yet unidentified intrinsic sources.} 
%such as the shot noise of the junction. 
%We plan to study the precise nature of the noise is unknown.

In conclusion, we have studied the AC Josephson effect in a non-topological graphene junction, which allows one to directly tune many of the relevant parameters. The type of samples studied here provides a highly tunable platform to probe the unexplored aspects of driven-dissipative dynamics of a quantum system. Understanding the variety of Shapiro patterns obtained in a prototypical graphene SNS junction will help to identify the non-trivial features in junctions made of topological materials~\cite{Rokhinson2012,Wiedenmann2016,Li4pi2018,Bocquillon2017}. It is also opens interesting perspectives for studying multi-terminal junctions~\cite{Strambini2016,AnneMultiterminal,MartinTopoFreqConv2017}, which could reveal topological bands when subject to RF drive~\cite{Gavensky2018}. Finally, since the Hamiltonian of the RF-driven junction is periodic in the phase difference and in time, it could be considered in the context of Floquet physics, which could potentially result in topologically non-trivial bands (see e.g. Ref.~\cite{HarperARP2020}). 

%\GF{Move to conclusion?} Furthermore, the phase space of multi-terminal junctions can have topological properties which could be revealed when subject to RF drive.~\cite{Klees} 

\begin{acknowledgements}
We thank M. Dykman, S. Teitsworth and J. Williams for helpful discussions.  Low-temperature electronic measurements performed by T.F.L and E.G.A, as well as simulations performed by L.Z., were supported by the Office of Basic Energy Sciences, U.S. Department of Energy, under Award DE-SC0002765. Lithographic fabrication and characterization of the samples performed by M.T.W. and A.S. were supported by ARO Award W911NF16-1- 0122. H. Li, and F.A. acknowledge the ARO under Award W911NF-16-1-0132.  K.W. and T.T. acknowledge the Elemental Strategy Initiative conducted by the MEXT, Japan and the CREST (JPMJCR15F3), JST. S.T. and M. Y. acknowledges KAKENHI (GrantNo. 38000131, 17H01138).  This work was performed in part at the Duke University Shared Materials Instrumentation Facility (SMIF), a member of the North Carolina Research Triangle Nanotechnology Network (RTNN), which is supported by the National Science Foundation (Grant ECCS-1542015) as part of the National Nanotechnology Coordinated Infrastructure (NNCI). 
\end{acknowledgements}

\bibliography{TrevynLibrary}

\begin{thebibliography}{32}
\expandafter\ifx\csname natexlab\endcsname\relax\def\natexlab#1{#1}\fi
\expandafter\ifx\csname bibnamefont\endcsname\relax
  \def\bibnamefont#1{#1}\fi
\expandafter\ifx\csname bibfnamefont\endcsname\relax
  \def\bibfnamefont#1{#1}\fi
\expandafter\ifx\csname citenamefont\endcsname\relax
  \def\citenamefont#1{#1}\fi
\expandafter\ifx\csname url\endcsname\relax
  \def\url#1{\texttt{#1}}\fi
\expandafter\ifx\csname urlprefix\endcsname\relax\def\urlprefix{URL }\fi
\providecommand{\bibinfo}[2]{#2}
\providecommand{\eprint}[2][]{\url{#2}}

\bibitem[{\citenamefont{Josephson}(1964)}]{JosephsonRMP1964}
\bibinfo{author}{\bibfnamefont{B.~D.} \bibnamefont{Josephson}},
  \bibinfo{journal}{Reviews of Modern Physics} \textbf{\bibinfo{volume}{36}},
  \bibinfo{pages}{216} (\bibinfo{year}{1964}), ISSN \bibinfo{issn}{0034-6861}.

\bibitem[{\citenamefont{Shapiro}(1963)}]{Shapiro1963}
\bibinfo{author}{\bibfnamefont{S.}~\bibnamefont{Shapiro}},
  \bibinfo{journal}{Physical Review Letters} \textbf{\bibinfo{volume}{11}},
  \bibinfo{pages}{80} (\bibinfo{year}{1963}), ISSN \bibinfo{issn}{0031-9007}.

\bibitem[{\citenamefont{Kautz}(1996)}]{KautzRev1996}
\bibinfo{author}{\bibfnamefont{R.~L.} \bibnamefont{Kautz}},
  \bibinfo{journal}{Reports on Progress in Physics}
  \textbf{\bibinfo{volume}{59}}, \bibinfo{pages}{935} (\bibinfo{year}{1996}),
  ISSN \bibinfo{issn}{0034-4885, 1361-6633}.

\bibitem[{\citenamefont{Hamilton}(2000)}]{hamiltonJosephsonVoltageStandards2000}
\bibinfo{author}{\bibfnamefont{C.~A.} \bibnamefont{Hamilton}},
  \bibinfo{journal}{Review of Scientific Instruments}
  \textbf{\bibinfo{volume}{71}} (\bibinfo{year}{2000}).

\bibitem[{\citenamefont{Kwon et~al.}(2004)\citenamefont{Kwon, Yakovenko, and
  Sengupta}}]{Kwon2004}
\bibinfo{author}{\bibfnamefont{H.-J.} \bibnamefont{Kwon}},
  \bibinfo{author}{\bibfnamefont{V.~M.} \bibnamefont{Yakovenko}},
  \bibnamefont{and} \bibinfo{author}{\bibfnamefont{K.}~\bibnamefont{Sengupta}},
  \bibinfo{journal}{Low Temperature Physics} \textbf{\bibinfo{volume}{30}},
  \bibinfo{pages}{613} (\bibinfo{year}{2004}), ISSN \bibinfo{issn}{1063-777X,
  1090-6517}.

\bibitem[{\citenamefont{Fu and Kane}(2009)}]{FuKane2009}
\bibinfo{author}{\bibfnamefont{L.}~\bibnamefont{Fu}} \bibnamefont{and}
  \bibinfo{author}{\bibfnamefont{C.~L.} \bibnamefont{Kane}},
  \bibinfo{journal}{Physical Review B} \textbf{\bibinfo{volume}{79}},
  \bibinfo{pages}{161408} (\bibinfo{year}{2009}), ISSN
  \bibinfo{issn}{1098-0121, 1550-235X}.

\bibitem[{\citenamefont{Rokhinson et~al.}(2012)\citenamefont{Rokhinson, Liu,
  and Furdyna}}]{Rokhinson2012}
\bibinfo{author}{\bibfnamefont{L.~P.} \bibnamefont{Rokhinson}},
  \bibinfo{author}{\bibfnamefont{X.}~\bibnamefont{Liu}}, \bibnamefont{and}
  \bibinfo{author}{\bibfnamefont{J.~K.} \bibnamefont{Furdyna}},
  \bibinfo{journal}{Nature Physics} \textbf{\bibinfo{volume}{8}},
  \bibinfo{pages}{795} (\bibinfo{year}{2012}), ISSN \bibinfo{issn}{1745-2473,
  1745-2481}.

\bibitem[{\citenamefont{Wiedenmann et~al.}(2016)\citenamefont{Wiedenmann,
  Bocquillon, Deacon, Hartinger, Herrmann, Klapwijk, Maier, Ames, Br{\"u}ne,
  Gould et~al.}}]{Wiedenmann2016}
\bibinfo{author}{\bibfnamefont{J.}~\bibnamefont{Wiedenmann}},
  \bibinfo{author}{\bibfnamefont{E.}~\bibnamefont{Bocquillon}},
  \bibinfo{author}{\bibfnamefont{R.~S.} \bibnamefont{Deacon}},
  \bibinfo{author}{\bibfnamefont{S.}~\bibnamefont{Hartinger}},
  \bibinfo{author}{\bibfnamefont{O.}~\bibnamefont{Herrmann}},
  \bibinfo{author}{\bibfnamefont{T.~M.} \bibnamefont{Klapwijk}},
  \bibinfo{author}{\bibfnamefont{L.}~\bibnamefont{Maier}},
  \bibinfo{author}{\bibfnamefont{C.}~\bibnamefont{Ames}},
  \bibinfo{author}{\bibfnamefont{C.}~\bibnamefont{Br{\"u}ne}},
  \bibinfo{author}{\bibfnamefont{C.}~\bibnamefont{Gould}},
  \bibnamefont{et~al.}, \bibinfo{journal}{Nature Communications}
  \textbf{\bibinfo{volume}{7}} (\bibinfo{year}{2016}), ISSN
  \bibinfo{issn}{2041-1723}.

\bibitem[{\citenamefont{Bocquillon et~al.}(2017)\citenamefont{Bocquillon,
  Deacon, Wiedenmann, Leubner, Klapwijk, Br{\"u}ne, Ishibashi, Buhmann, and
  Molenkamp}}]{Bocquillon2017}
\bibinfo{author}{\bibfnamefont{E.}~\bibnamefont{Bocquillon}},
  \bibinfo{author}{\bibfnamefont{R.~S.} \bibnamefont{Deacon}},
  \bibinfo{author}{\bibfnamefont{J.}~\bibnamefont{Wiedenmann}},
  \bibinfo{author}{\bibfnamefont{P.}~\bibnamefont{Leubner}},
  \bibinfo{author}{\bibfnamefont{T.~M.} \bibnamefont{Klapwijk}},
  \bibinfo{author}{\bibfnamefont{C.}~\bibnamefont{Br{\"u}ne}},
  \bibinfo{author}{\bibfnamefont{K.}~\bibnamefont{Ishibashi}},
  \bibinfo{author}{\bibfnamefont{H.}~\bibnamefont{Buhmann}}, \bibnamefont{and}
  \bibinfo{author}{\bibfnamefont{L.~W.} \bibnamefont{Molenkamp}},
  \bibinfo{journal}{Nature Nanotechnology} \textbf{\bibinfo{volume}{12}},
  \bibinfo{pages}{137} (\bibinfo{year}{2017}), ISSN \bibinfo{issn}{1748-3387,
  1748-3395}.

\bibitem[{\citenamefont{Li et~al.}(2018)\citenamefont{Li, {de Boer}, {de
  Ronde}, Ramankutty, {van Heumen}, Huang, {de Visser}, Golubov, Golden, and
  Brinkman}}]{Li4pi2018}
\bibinfo{author}{\bibfnamefont{C.}~\bibnamefont{Li}},
  \bibinfo{author}{\bibfnamefont{J.~C.} \bibnamefont{{de Boer}}},
  \bibinfo{author}{\bibfnamefont{B.}~\bibnamefont{{de Ronde}}},
  \bibinfo{author}{\bibfnamefont{S.~V.} \bibnamefont{Ramankutty}},
  \bibinfo{author}{\bibfnamefont{E.}~\bibnamefont{{van Heumen}}},
  \bibinfo{author}{\bibfnamefont{Y.}~\bibnamefont{Huang}},
  \bibinfo{author}{\bibfnamefont{A.}~\bibnamefont{{de Visser}}},
  \bibinfo{author}{\bibfnamefont{A.~A.} \bibnamefont{Golubov}},
  \bibinfo{author}{\bibfnamefont{M.~S.} \bibnamefont{Golden}},
  \bibnamefont{and} \bibinfo{author}{\bibfnamefont{A.}~\bibnamefont{Brinkman}},
  \bibinfo{journal}{Nature Materials} \textbf{\bibinfo{volume}{17}},
  \bibinfo{pages}{875} (\bibinfo{year}{2018}), ISSN \bibinfo{issn}{1476-1122,
  1476-4660}.

\bibitem[{\citenamefont{Le~Calvez et~al.}(2019)\citenamefont{Le~Calvez, Veyrat,
  Gay, Plaindoux, Winkelmann, Courtois, and Sac{\'e}p{\'e}}}]{Calvez2019}
\bibinfo{author}{\bibfnamefont{K.}~\bibnamefont{Le~Calvez}},
  \bibinfo{author}{\bibfnamefont{L.}~\bibnamefont{Veyrat}},
  \bibinfo{author}{\bibfnamefont{F.}~\bibnamefont{Gay}},
  \bibinfo{author}{\bibfnamefont{P.}~\bibnamefont{Plaindoux}},
  \bibinfo{author}{\bibfnamefont{C.~B.} \bibnamefont{Winkelmann}},
  \bibinfo{author}{\bibfnamefont{H.}~\bibnamefont{Courtois}}, \bibnamefont{and}
  \bibinfo{author}{\bibfnamefont{B.}~\bibnamefont{Sac{\'e}p{\'e}}},
  \bibinfo{journal}{Communications Physics} \textbf{\bibinfo{volume}{2}}
  (\bibinfo{year}{2019}), ISSN \bibinfo{issn}{2399-3650}.

\bibitem[{\citenamefont{Heersche et~al.}(2007)\citenamefont{Heersche,
  {Jarillo-Herrero}, Oostinga, Vandersypen, and Morpurgo}}]{Heersche2007}
\bibinfo{author}{\bibfnamefont{H.~B.} \bibnamefont{Heersche}},
  \bibinfo{author}{\bibfnamefont{P.}~\bibnamefont{{Jarillo-Herrero}}},
  \bibinfo{author}{\bibfnamefont{J.~B.} \bibnamefont{Oostinga}},
  \bibinfo{author}{\bibfnamefont{L.~M.~K.} \bibnamefont{Vandersypen}},
  \bibnamefont{and} \bibinfo{author}{\bibfnamefont{A.~F.}
  \bibnamefont{Morpurgo}}, \bibinfo{journal}{Nature}
  \textbf{\bibinfo{volume}{446}}, \bibinfo{pages}{56} (\bibinfo{year}{2007}),
  ISSN \bibinfo{issn}{0028-0836, 1476-4687}.

\bibitem[{\citenamefont{Komatsu et~al.}(2012)\citenamefont{Komatsu, Li,
  {Autier-Laurent}, Bouchiat, and Gu{\'e}ron}}]{Komatsu2012}
\bibinfo{author}{\bibfnamefont{K.}~\bibnamefont{Komatsu}},
  \bibinfo{author}{\bibfnamefont{C.}~\bibnamefont{Li}},
  \bibinfo{author}{\bibfnamefont{S.}~\bibnamefont{{Autier-Laurent}}},
  \bibinfo{author}{\bibfnamefont{H.}~\bibnamefont{Bouchiat}}, \bibnamefont{and}
  \bibinfo{author}{\bibfnamefont{S.}~\bibnamefont{Gu{\'e}ron}},
  \bibinfo{journal}{Physical Review B} \textbf{\bibinfo{volume}{86}},
  \bibinfo{pages}{115412} (\bibinfo{year}{2012}), ISSN
  \bibinfo{issn}{1098-0121, 1550-235X}.

\bibitem[{\citenamefont{Lee et~al.}(2015)\citenamefont{Lee, Kim, Jhi, and
  Lee}}]{LeeVertJunction2015}
\bibinfo{author}{\bibfnamefont{G.-H.} \bibnamefont{Lee}},
  \bibinfo{author}{\bibfnamefont{S.}~\bibnamefont{Kim}},
  \bibinfo{author}{\bibfnamefont{S.-H.} \bibnamefont{Jhi}}, \bibnamefont{and}
  \bibinfo{author}{\bibfnamefont{H.-J.} \bibnamefont{Lee}},
  \bibinfo{journal}{Nature Communications} \textbf{\bibinfo{volume}{6}},
  \bibinfo{pages}{6181} (\bibinfo{year}{2015}), ISSN \bibinfo{issn}{2041-1723}.

\bibitem[{\citenamefont{{Kalantre} et~al.}(2019)\citenamefont{{Kalantre}, {Yu},
  {Wei}, {Watanabe}, {Taniguchi}, {Hernandez-Rivera}, {Amet}, and
  {Williams}}}]{kalantreAnomalousPhaseDynamics2019}
\bibinfo{author}{\bibfnamefont{S.~S.} \bibnamefont{{Kalantre}}},
  \bibinfo{author}{\bibfnamefont{F.}~\bibnamefont{{Yu}}},
  \bibinfo{author}{\bibfnamefont{M.~T.} \bibnamefont{{Wei}}},
  \bibinfo{author}{\bibfnamefont{K.}~\bibnamefont{{Watanabe}}},
  \bibinfo{author}{\bibfnamefont{T.}~\bibnamefont{{Taniguchi}}},
  \bibinfo{author}{\bibfnamefont{M.}~\bibnamefont{{Hernandez-Rivera}}},
  \bibinfo{author}{\bibfnamefont{F.}~\bibnamefont{{Amet}}}, \bibnamefont{and}
  \bibinfo{author}{\bibfnamefont{J.~R.} \bibnamefont{{Williams}}},
  \bibinfo{journal}{arXiv} \bibinfo{eid}{arXiv:1910.10125}
  (\bibinfo{year}{2019}), \eprint{1910.10125}.

\bibitem[{Wie()}]{WiedenmannSupp}
\bibinfo{note}{See supplemental of Ref.[8]}.

\bibitem[{\citenamefont{Dean et~al.}(2010)\citenamefont{Dean, Young, Meric,
  Lee, Wang, Sorgenfrei, Watanabe, Taniguchi, Kim, Shepard et~al.}}]{Dean2010}
\bibinfo{author}{\bibfnamefont{C.~R.} \bibnamefont{Dean}},
  \bibinfo{author}{\bibfnamefont{A.~F.} \bibnamefont{Young}},
  \bibinfo{author}{\bibfnamefont{I.}~\bibnamefont{Meric}},
  \bibinfo{author}{\bibfnamefont{C.}~\bibnamefont{Lee}},
  \bibinfo{author}{\bibfnamefont{L.}~\bibnamefont{Wang}},
  \bibinfo{author}{\bibfnamefont{S.}~\bibnamefont{Sorgenfrei}},
  \bibinfo{author}{\bibfnamefont{K.}~\bibnamefont{Watanabe}},
  \bibinfo{author}{\bibfnamefont{T.}~\bibnamefont{Taniguchi}},
  \bibinfo{author}{\bibfnamefont{P.}~\bibnamefont{Kim}},
  \bibinfo{author}{\bibfnamefont{K.~L.} \bibnamefont{Shepard}},
  \bibnamefont{et~al.}, \bibinfo{journal}{Nature Nanotechnology}
  \textbf{\bibinfo{volume}{5}}, \bibinfo{pages}{722} (\bibinfo{year}{2010}),
  ISSN \bibinfo{issn}{1748-3387, 1748-3395}.

\bibitem[{\citenamefont{Calado et~al.}(2015)\citenamefont{Calado, Goswami,
  Nanda, Diez, Akhmerov, Watanabe, Taniguchi, Klapwijk, and
  Vandersypen}}]{CaladoMoRe}
\bibinfo{author}{\bibfnamefont{V.~E.} \bibnamefont{Calado}},
  \bibinfo{author}{\bibfnamefont{S.}~\bibnamefont{Goswami}},
  \bibinfo{author}{\bibfnamefont{G.}~\bibnamefont{Nanda}},
  \bibinfo{author}{\bibfnamefont{M.}~\bibnamefont{Diez}},
  \bibinfo{author}{\bibfnamefont{A.~R.} \bibnamefont{Akhmerov}},
  \bibinfo{author}{\bibfnamefont{K.}~\bibnamefont{Watanabe}},
  \bibinfo{author}{\bibfnamefont{T.}~\bibnamefont{Taniguchi}},
  \bibinfo{author}{\bibfnamefont{T.~M.} \bibnamefont{Klapwijk}},
  \bibnamefont{and} \bibinfo{author}{\bibfnamefont{L.~M.~K.}
  \bibnamefont{Vandersypen}}, \bibinfo{journal}{Nature Nanotechnology}
  \textbf{\bibinfo{volume}{10}}, \bibinfo{pages}{761} (\bibinfo{year}{2015}),
  ISSN \bibinfo{issn}{1748-3387, 1748-3395}.

\bibitem[{\citenamefont{De~Cecco et~al.}(2016)\citenamefont{De~Cecco,
  Le~Calvez, Sac{\'e}p{\'e}, Winkelmann, and Courtois}}]{DeCecco2016}
\bibinfo{author}{\bibfnamefont{A.}~\bibnamefont{De~Cecco}},
  \bibinfo{author}{\bibfnamefont{K.}~\bibnamefont{Le~Calvez}},
  \bibinfo{author}{\bibfnamefont{B.}~\bibnamefont{Sac{\'e}p{\'e}}},
  \bibinfo{author}{\bibfnamefont{C.~B.} \bibnamefont{Winkelmann}},
  \bibnamefont{and} \bibinfo{author}{\bibfnamefont{H.}~\bibnamefont{Courtois}},
  \bibinfo{journal}{Physical Review B} \textbf{\bibinfo{volume}{93}}
  (\bibinfo{year}{2016}), ISSN \bibinfo{issn}{2469-9950, 2469-9969}.

\bibitem[{\citenamefont{Tinkham}(1996)}]{Tinkham}
\bibinfo{author}{\bibfnamefont{M.}~\bibnamefont{Tinkham}},
  \emph{\bibinfo{title}{Introduction to {{Superconductivity}}}}
  (\bibinfo{publisher}{{Dover}}, \bibinfo{year}{1996}), \bibinfo{edition}{2nd}
  ed.

\bibitem[{\citenamefont{Russer}(1972)}]{Russer1972}
\bibinfo{author}{\bibfnamefont{P.}~\bibnamefont{Russer}},
  \bibinfo{journal}{Journal of Applied Physics} \textbf{\bibinfo{volume}{43}},
  \bibinfo{pages}{2008} (\bibinfo{year}{1972}), ISSN \bibinfo{issn}{0021-8979,
  1089-7550}.

\bibitem[{\citenamefont{Thorne et~al.}(1987)\citenamefont{Thorne, Lyons,
  Lyding, Tucker, and Bardeen}}]{ThorneBardeenpt2}
\bibinfo{author}{\bibfnamefont{R.~E.} \bibnamefont{Thorne}},
  \bibinfo{author}{\bibfnamefont{W.~G.} \bibnamefont{Lyons}},
  \bibinfo{author}{\bibfnamefont{J.~W.} \bibnamefont{Lyding}},
  \bibinfo{author}{\bibfnamefont{J.~R.} \bibnamefont{Tucker}},
  \bibnamefont{and} \bibinfo{author}{\bibfnamefont{J.}~\bibnamefont{Bardeen}},
  \bibinfo{journal}{Physical Review B} \textbf{\bibinfo{volume}{35}},
  \bibinfo{pages}{6360} (\bibinfo{year}{1987}), ISSN \bibinfo{issn}{0163-1829}.

\bibitem[{\citenamefont{Strambini et~al.}(2016)\citenamefont{Strambini,
  D'Ambrosio, Vischi, Bergeret, Nazarov, and Giazotto}}]{Strambini2016}
\bibinfo{author}{\bibfnamefont{E.}~\bibnamefont{Strambini}},
  \bibinfo{author}{\bibfnamefont{S.}~\bibnamefont{D'Ambrosio}},
  \bibinfo{author}{\bibfnamefont{F.}~\bibnamefont{Vischi}},
  \bibinfo{author}{\bibfnamefont{F.~S.} \bibnamefont{Bergeret}},
  \bibinfo{author}{\bibfnamefont{Y.~V.} \bibnamefont{Nazarov}},
  \bibnamefont{and} \bibinfo{author}{\bibfnamefont{F.}~\bibnamefont{Giazotto}},
  \bibinfo{journal}{Nature Nanotechnology} \textbf{\bibinfo{volume}{11}},
  \bibinfo{pages}{1055} (\bibinfo{year}{2016}), ISSN \bibinfo{issn}{1748-3387,
  1748-3395}.

\bibitem[{\citenamefont{Draelos et~al.}(2019)\citenamefont{Draelos, Wei,
  Seredinski, Li, Mehta, Watanabe, Taniguchi, Borzenets, Amet, and
  Finkelstein}}]{AnneMultiterminal}
\bibinfo{author}{\bibfnamefont{A.~W.} \bibnamefont{Draelos}},
  \bibinfo{author}{\bibfnamefont{M.-T.} \bibnamefont{Wei}},
  \bibinfo{author}{\bibfnamefont{A.}~\bibnamefont{Seredinski}},
  \bibinfo{author}{\bibfnamefont{H.}~\bibnamefont{Li}},
  \bibinfo{author}{\bibfnamefont{Y.}~\bibnamefont{Mehta}},
  \bibinfo{author}{\bibfnamefont{K.}~\bibnamefont{Watanabe}},
  \bibinfo{author}{\bibfnamefont{T.}~\bibnamefont{Taniguchi}},
  \bibinfo{author}{\bibfnamefont{I.~V.} \bibnamefont{Borzenets}},
  \bibinfo{author}{\bibfnamefont{F.}~\bibnamefont{Amet}}, \bibnamefont{and}
  \bibinfo{author}{\bibfnamefont{G.}~\bibnamefont{Finkelstein}},
  \bibinfo{journal}{Nano Letters} \textbf{\bibinfo{volume}{19}},
  \bibinfo{pages}{1039} (\bibinfo{year}{2019}), ISSN \bibinfo{issn}{1530-6984,
  1530-6992}.

\bibitem[{\citenamefont{Martin et~al.}(2017)\citenamefont{Martin, Refael, and
  Halperin}}]{MartinTopoFreqConv2017}
\bibinfo{author}{\bibfnamefont{I.}~\bibnamefont{Martin}},
  \bibinfo{author}{\bibfnamefont{G.}~\bibnamefont{Refael}}, \bibnamefont{and}
  \bibinfo{author}{\bibfnamefont{B.}~\bibnamefont{Halperin}},
  \bibinfo{journal}{Physical Review X} \textbf{\bibinfo{volume}{7}},
  \bibinfo{pages}{041008} (\bibinfo{year}{2017}), ISSN
  \bibinfo{issn}{2160-3308}.

\bibitem[{\citenamefont{Peralta~Gavensky
  et~al.}(2018)\citenamefont{Peralta~Gavensky, Usaj, Feinberg, and
  Balseiro}}]{Gavensky2018}
\bibinfo{author}{\bibfnamefont{L.}~\bibnamefont{Peralta~Gavensky}},
  \bibinfo{author}{\bibfnamefont{G.}~\bibnamefont{Usaj}},
  \bibinfo{author}{\bibfnamefont{D.}~\bibnamefont{Feinberg}}, \bibnamefont{and}
  \bibinfo{author}{\bibfnamefont{C.~A.} \bibnamefont{Balseiro}},
  \bibinfo{journal}{Physical Review B} \textbf{\bibinfo{volume}{97}},
  \bibinfo{pages}{220505} (\bibinfo{year}{2018}), ISSN
  \bibinfo{issn}{2469-9950, 2469-9969}.

\bibitem[{\citenamefont{Harper et~al.}(2020)\citenamefont{Harper, Roy, Rudner,
  and Sondhi}}]{HarperARP2020}
\bibinfo{author}{\bibfnamefont{F.}~\bibnamefont{Harper}},
  \bibinfo{author}{\bibfnamefont{R.}~\bibnamefont{Roy}},
  \bibinfo{author}{\bibfnamefont{M.~S.} \bibnamefont{Rudner}},
  \bibnamefont{and} \bibinfo{author}{\bibfnamefont{S.}~\bibnamefont{Sondhi}},
  \bibinfo{journal}{Annual Review of Condensed Matter Physics}
  \textbf{\bibinfo{volume}{11}} (\bibinfo{year}{2020}), ISSN
  \bibinfo{issn}{1947-5454, 1947-5462}.

\bibitem[{\citenamefont{Clarke et~al.}(1988)\citenamefont{Clarke, Cleland,
  Devoret, Esteve, and Martinis}}]{Clarke1988}
\bibinfo{author}{\bibfnamefont{J.}~\bibnamefont{Clarke}},
  \bibinfo{author}{\bibfnamefont{A.~N.} \bibnamefont{Cleland}},
  \bibinfo{author}{\bibfnamefont{M.~H.} \bibnamefont{Devoret}},
  \bibinfo{author}{\bibfnamefont{D.}~\bibnamefont{Esteve}}, \bibnamefont{and}
  \bibinfo{author}{\bibfnamefont{J.~M.} \bibnamefont{Martinis}},
  \bibinfo{journal}{Science} \textbf{\bibinfo{volume}{239}},
  \bibinfo{pages}{992} (\bibinfo{year}{1988}), ISSN \bibinfo{issn}{0036-8075,
  1095-9203}.

\bibitem[{\citenamefont{{Jarillo-Herrero}
  et~al.}(2006)\citenamefont{{Jarillo-Herrero}, {van Dam}, and
  Kouwenhoven}}]{jarillo-herreroQuantumSupercurrentTransistors2006}
\bibinfo{author}{\bibfnamefont{P.}~\bibnamefont{{Jarillo-Herrero}}},
  \bibinfo{author}{\bibfnamefont{J.~A.} \bibnamefont{{van Dam}}},
  \bibnamefont{and} \bibinfo{author}{\bibfnamefont{L.~P.}
  \bibnamefont{Kouwenhoven}}, \bibinfo{journal}{Nature}
  \textbf{\bibinfo{volume}{439}}, \bibinfo{pages}{953} (\bibinfo{year}{2006}),
  ISSN \bibinfo{issn}{0028-0836, 1476-4687}.

\bibitem[{\citenamefont{Anderson and Dayem}(1964)}]{AndersonDayem1964}
\bibinfo{author}{\bibfnamefont{P.~W.} \bibnamefont{Anderson}} \bibnamefont{and}
  \bibinfo{author}{\bibfnamefont{A.~H.} \bibnamefont{Dayem}},
  \bibinfo{journal}{Physical Review Letters} \textbf{\bibinfo{volume}{13}},
  \bibinfo{pages}{195} (\bibinfo{year}{1964}), ISSN \bibinfo{issn}{0031-9007}.

\bibitem[{\citenamefont{English et~al.}(2016)\citenamefont{English, Hamilton,
  Chialvo, Moraru, Mason, and Van~Harlingen}}]{EnglishCPR2016}
\bibinfo{author}{\bibfnamefont{C.~D.} \bibnamefont{English}},
  \bibinfo{author}{\bibfnamefont{D.~R.} \bibnamefont{Hamilton}},
  \bibinfo{author}{\bibfnamefont{C.}~\bibnamefont{Chialvo}},
  \bibinfo{author}{\bibfnamefont{I.~C.} \bibnamefont{Moraru}},
  \bibinfo{author}{\bibfnamefont{N.}~\bibnamefont{Mason}}, \bibnamefont{and}
  \bibinfo{author}{\bibfnamefont{D.~J.} \bibnamefont{Van~Harlingen}},
  \bibinfo{journal}{Physical Review B} \textbf{\bibinfo{volume}{94}},
  \bibinfo{pages}{115435} (\bibinfo{year}{2016}), ISSN
  \bibinfo{issn}{2469-9950, 2469-9969}.

\bibitem[{\citenamefont{Nanda et~al.}(2017)\citenamefont{Nanda,
  {Aguilera-Servin}, Rakyta, Korm{\'a}nyos, Kleiner, Koelle, Watanabe,
  Taniguchi, Vandersypen, and Goswami}}]{Nanda2017}
\bibinfo{author}{\bibfnamefont{G.}~\bibnamefont{Nanda}},
  \bibinfo{author}{\bibfnamefont{J.~L.} \bibnamefont{{Aguilera-Servin}}},
  \bibinfo{author}{\bibfnamefont{P.}~\bibnamefont{Rakyta}},
  \bibinfo{author}{\bibfnamefont{A.}~\bibnamefont{Korm{\'a}nyos}},
  \bibinfo{author}{\bibfnamefont{R.}~\bibnamefont{Kleiner}},
  \bibinfo{author}{\bibfnamefont{D.}~\bibnamefont{Koelle}},
  \bibinfo{author}{\bibfnamefont{K.}~\bibnamefont{Watanabe}},
  \bibinfo{author}{\bibfnamefont{T.}~\bibnamefont{Taniguchi}},
  \bibinfo{author}{\bibfnamefont{L.~M.~K.} \bibnamefont{Vandersypen}},
  \bibnamefont{and} \bibinfo{author}{\bibfnamefont{S.}~\bibnamefont{Goswami}},
  \bibinfo{journal}{Nano Letters} \textbf{\bibinfo{volume}{17}},
  \bibinfo{pages}{3396} (\bibinfo{year}{2017}), ISSN \bibinfo{issn}{1530-6984,
  1530-6992}.

\end{thebibliography}
\bibliographystyle{apsrev}

\newpage

\clearpage

\begin{center}
\textbf{\large Supplemental Materials}
\end{center}
\renewcommand{\thepage}{S\arabic{page}}
\renewcommand{\thefigure}{S\arabic{figure}}
\renewcommand{\thesection}{S\arabic{section}}
\setcounter{figure}{0}
\setcounter{page}{1}
\setcounter{equation}{0}

\section{Sample Information}
The  hexagonal boron nitride/ graphene heterostructure was assembled by the standard dry transfer stamping method from mechanically exfoliated flakes. The stack was then deposited on a silicon chip with a 280 nm oxide layer. The contacts were made of sputtered Molybdenum-Rhenium alloy (50-50 by weight), a type-II superconductor with a high critical temperature of $8-10$ K. The junction has dimensions of 0.5 x 3$\mu m$ and has been previously used as a reference device in Ref.~\onlinecite{AnneMultiterminal}. The MoRe contacts were connected to measurement lines through Cr/Au leads and bonding pads. By comparison to the simulations, we believe these leads and bonding pads act as the part of the environment which determines junction dynamics~\cite{Clarke1988}. We model them as a resistor $R_L$ in series with the capacitor $C_0$. 

\section{Measurement Techniques}
While differential resistance maps are often measured using a lock-in amplifier, in this case hysteretic switching prevented us from using this technique. Instead, for each vertical line of the maps, an approximately 20 Hz triangle wave was applied and the resulting voltage profile was measured. 200 such measurements were then averaged to produce an $I-V$ curve and then numerically differentiated. These parameters allowed for fast measurements with reasonable averaging and minimal distortion of the applied wave by the low-temperature filters. We note that the exact extent of hysteresis is a function of the sweeping speed, with faster sweeping giving more pronounced hysteresis.  This is particularly relevant when comparing to the numerical results, as the simulations used much shorter time evolutions thereby exaggerating the effects of the hysteresis.

\section{Simulations}

\begin{figure}[htb]
\centering
\includegraphics[width=0.35\textwidth]{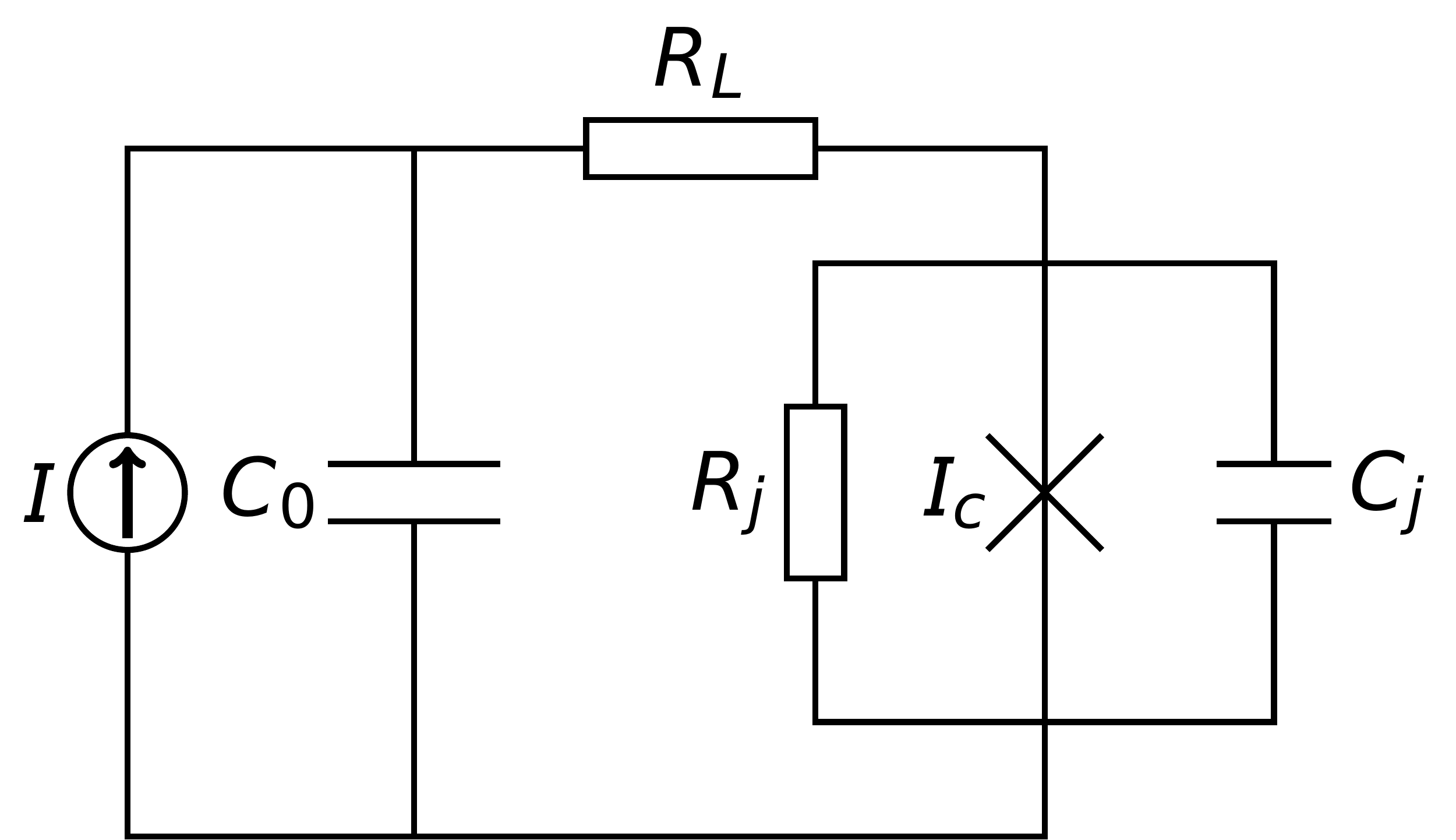}
\caption{Diagram of the circuit used to simulate the dynamics of the Josephson junction. In practice, $C_{j}$ is negligible and is omitted from further consideration.}
\label{circuit}
\end{figure}

To simulate the behavior of a Josephson junction subject to microwave radiation, we use a modified RCSJ model as illustrated in Fig.~\ref{circuit}~\cite{jarillo-herreroQuantumSupercurrentTransistors2006}. We start with a junction with critical current $I_C$, which is shunted by a resistor $R_j$ and a capacitor $C_j$, where $R_j$ represents the dissipation in the Josephson junction and $C_j$ is the capacitance between the two superconducting leads. In the experiment, the Josephson junction is further connected to four 150 $\mu$m $\times$ 100 $\mu$m bonding pads by Cr/Au leads. The capacitance of the bonding pads, $C_0$, and the resistance of the leads, $R_L$, must be taken into account to properly simulate the junction dynamics. The four bonding pads are arranged such that the effective capacitance is equal to that of one bonding pad to the back gate, which would yield 1.8 pF for 280 nm thick SiO$_{2}$. At room temperature, the capacitance between two bonding pads and bonding wires connected to the chip carrier by bonding wire was measured to be slightly higher, around 2.5pF, which was the value used in the simulations. In practice, similar maps have been simulated using a range of $C_0$ values. 

The resistance of the evaporated Cr (5 nm)/Au (45 nm) film was measured to be 0.5 Ohm/$\msquare$, from which we estimate that $R_L$ is a few tens of Ohms for our typical devices. We use a reasonable value of $R_L=50$ Ohms for our simulations in Figure 3 of the main text and in the simulations below. Finally, $R_j=300$ Ohms is determined from the current corresponding to the center of the Shapiro plateaus in the Bessel function regime, $I_n=n\hbar \omega /R_j$. In accordance with the experiment, we assume that $R_j$ does not depend on magnetic field. The same value of $R_j=300$ Ohms is used to simulate all panels in Figure S2.

\begin{figure*}[ht!]
\centering
\includegraphics[width=\textwidth]{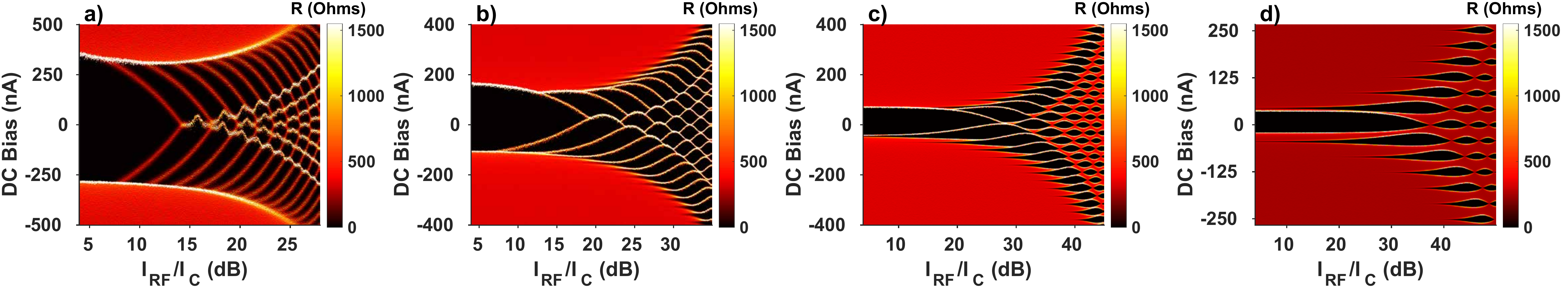}
    \caption{Simulations of the differential resistance maps at 5 GHz for comparison to Figure 1a-d. The values of $I_{C}$ used are, from left to right, 540, 200, 80 and 40 nA. Other parameters are kept the same as in Figure 3 of the main text: $C_{0}=2.5$ pF, $R_L=50$ Ohms and $R_j=300$ Ohms. }
\label{FrequencySims}
\end{figure*}

The microwave injection from the antenna can be modeled by a AC current, $I_{AC}=I_{RF} \sin{\omega t}$ where $I_{RF}$ is the current amplitude and $\omega$ is the microwave frequency. Note that a significant amount of the applied power is dropped across the capacitors in our model. The thermal noise of the resistive components in the experiment generates a Gaussian current noise $I_N$ whose variance is proportional to the temperature $T$. We find that for good agreement with the data, the noise amplitude has to be taken higher than expected for thermal noise. This is expected, for two reasons: first, in simulations $I_N$ is applied to the outside of the junction circuit, where it would be partially filtered by $C_0$ and $R_L$ before reaching the junction. Second, while each point on the map is measured for millions of cycles, we simulate it over just $\sim 500$ drive cycles, so using a higher level of noise may be expected. 
%Since higher noise accelerates the escape from a metastable state, we assume that higher noise applied over a shorter time is effectively similar 
We therefore, treat $I_N$ as a fitting parameter. In summary, the current source $I$ contains three components, the bias current, $I_{bias}$, the microwave radiation current, $I_{AC}$ and the thermal noise, $I_N$.

\begin{equation}
    \begin{aligned}
        I&=I_{bias}+I_{RF} \sin{\omega t}+I_N(t)\\
        &=C_0 \frac{dV}{dt}+ I_c \sin{\phi} +\frac{\hbar}{2 e R_j}\frac{d\phi}{dt}+\frac{\hbar C_j}{2 e}\frac{d^2 \phi}{dt^2}\\
        V&=\frac{\hbar}{2e}\frac{d\phi}{dt}+R_L\left(I_c \sin{\phi} +\frac{\hbar}{2 e R_j}\frac{d\phi}{dt}+\frac{\hbar C_j}{2 e}\frac{d^2 \phi}{dt^2}\right)
    \end{aligned}
    \label{RCSJ:eq1}
\end{equation}

The dynamics of the circuit in Fig.~\ref{circuit} is described by Eq.~(\ref{RCSJ:eq1}), where $\phi$ is the superconducting phase difference across the junction, $V$ is the voltage across the capacitor $C_0$ and $I_{N} \propto \sqrt{T}$ is the standard deviation of the Gaussian noise. Solving this third order differential equation numerically gives $\phi(t)$, from which we can derive the DC voltage across the junction, $V_j=\left<\frac{\hbar}{2e}\frac{d\phi}{dt}\right>$. Note that $C_j$ is about 4 orders of magnitude smaller than $C_0$ for the device studied here. We numerically found that $C_j$ can be neglected under this condition, simplifying Eq.~(\ref{RCSJ:eq1}) to a second order differential equation. The experimental curves strongly depend on the bias sweeping direction. To emulate the bias sweep, we use the steady solution of $\phi(t)$ at a given $I_{bias}$ as the initial condition for solving the differential equation at the next value of bias, $I_{bias}+\delta I$, where $\delta I$ is the incremental bias step.

%\section{Simulation of $I_{c}$ Dependence}

Figure S2 shows simulated Shapiro maps at several values of the critical current, intended to be compared with the 5 GHz data of Figure 1.  Remarkably, we are able to reproduce the four experimental maps in Figure 1a-d by changing only $I_{C}$, which is the only parameter we expect to be influenced by magnetic field. The values of $C_{0}=2.5$ pF, $R_L=50$ Ohms, and $R_j=300$ Ohms are kept the same in all panels.
%some liberty is also taken with the exact values of $I_{c}$. 

%We attribute the fractional steps present in Figure 1(a) and (d) to the skewed CPR expected for ballistic junctions. This effect has been left out of the simulation but can easily be incorporated. 
Our model does not include heating from the RF drive, which has been been recently identified to be an important effect~\cite{DeCecco2016}.  We believe that such heating offers a natural explanation for the washed out features at higher RF power observed in the experimental data.  %\TL{Should we discuss possible reasons for failing to match the envelope super well in simulations?  Heating, microwave enhanced gap, short coming of current bias assumption?} 
Overall, the simple model describes the main features of our data sufficiently well.

%Figure 1a also contains a fractional step which we attribute to the slightly skewed CPR of the junction. 

%This factor can naturally be incorporated into the simulations, but shows minimal effect at higher values of $I_{c}$ for any reasonable skewness.

\section{Simulation of Gate Voltage dependence}
Next, we reproduce the gate voltage dependence measured in Figure 4. Between the four maps, we adjusted the values of $I_{c}$ and $R_{j}$, where the former can be obtained from the value of the switching current at zero RF power, and the latter could be roughly extracted from the positions of the Shapiro steps, $I_n=n\hbar \omega /R_J$. Additionally, the noise amplitude $I_{N}$ was taken to be $ \propto 1/\sqrt{R_j}$. This is consistent with the expectation that the noise in the junction is given by the thermal noise of $R_{j}$.  While it may be expected that the noise processes in our driven system may be more complicated, the simulations capture the general trends observed in the data of Figure 4 of the main text.

%\GF{Why not take $I_{N} \propto 1/\sqrt{R_j}$?}

 %\TL{No good citation here.  Lots of early 2000s stuff on diffusive SNS e.g. https://www.sciencedirect.com/science/article/pii/S0921453400016762 Should we look to measure this?}. 

\begin{figure}[h!]
\centering
\includegraphics[width = 0.45\textwidth]{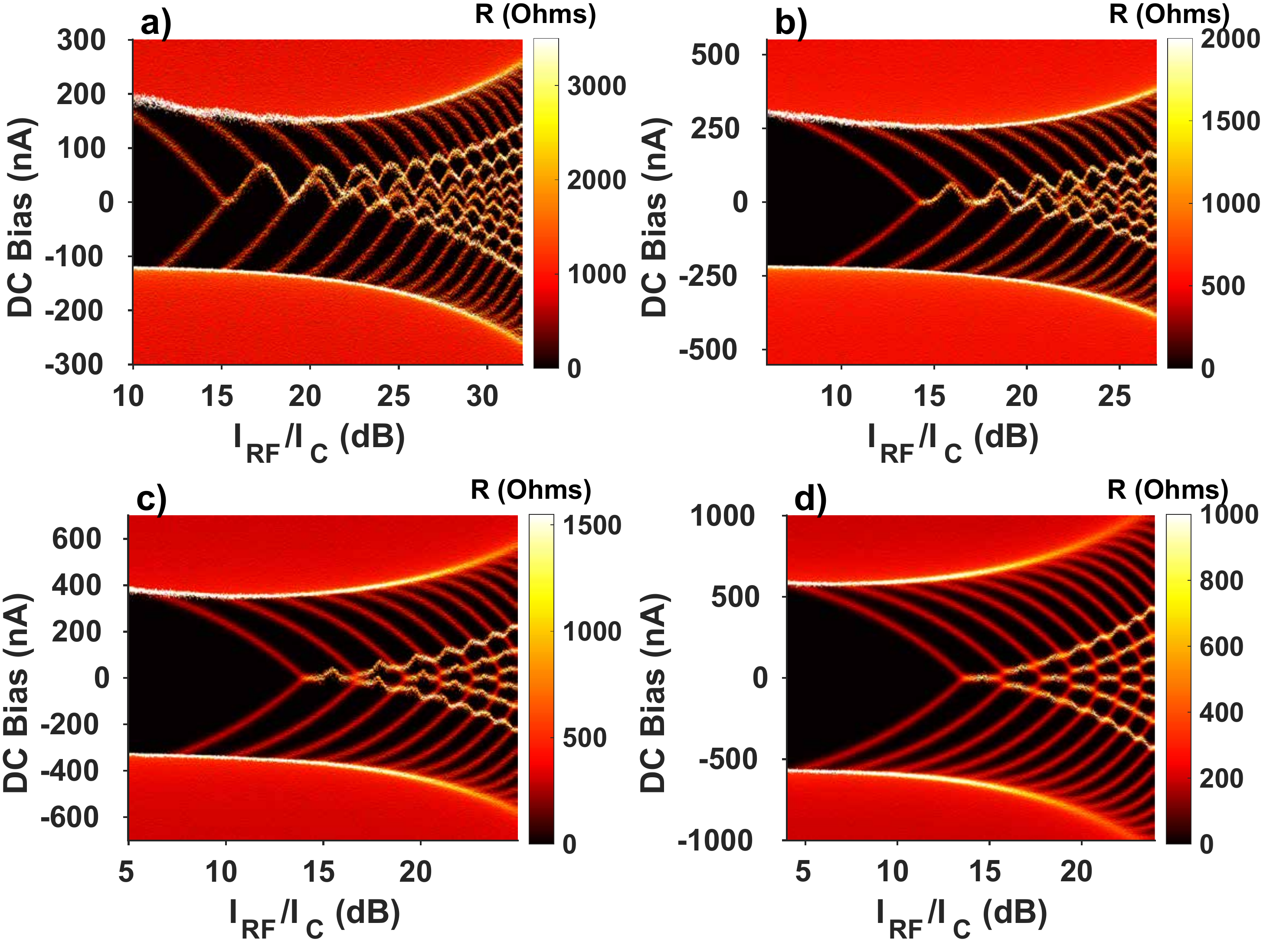}
    \caption{Simulations of differential resistance maps corresponding to Figure 4 of the main text, measured at different gate voltages. We use the values of $R_{j} =850,500,300,180$ Ohms and $I_{c} =350,500,600,800$ nA in panels (a) to (d).}
\label{GateSims}
\end{figure}
%\section{Hysteresis as a Function of RF Frequency}

%\begin{figure}[h!]
%\centering
%\includegraphics[width=0.3\textwidth]{QuasipotFigure.pdf}
%\caption{Symmetrized Shapiro step maps, where measurements using both sweep directions are summed together to emphasize hysteresis. The parameters are 0.45 $V_{G}$ and 5 GHz, 0 $V_{G}$ and 5 GHz and 0 $V_{G}$ and 6.4 GHz for a, b and c respectively.}
%\label{QuasipotFigure}
%\end{figure}

%The sweep direction dependence of the measurement is further emphasized in Figure S1a. Here, we fix the same conditions as in Figure 2a of the main text and sum the differential resistance measured in both sweep directions. This procedure emphasizes the fact that the boundaries in region II (the center of the map, see Figure 2b) are hysteretic. In Figure S1c, we increase the drive frequency to 6.4 GHz resulting in a large hysteresis now visible in both regions I and II. This change is likely explained by the increased ``quasipotential'',~\cite{KautzRev1996} which the system is required to overcome when switching between different Shapiro steps, and the corresponding slowing of the switching rates.

\begin{figure*}
\centering
\includegraphics[width = \textwidth]{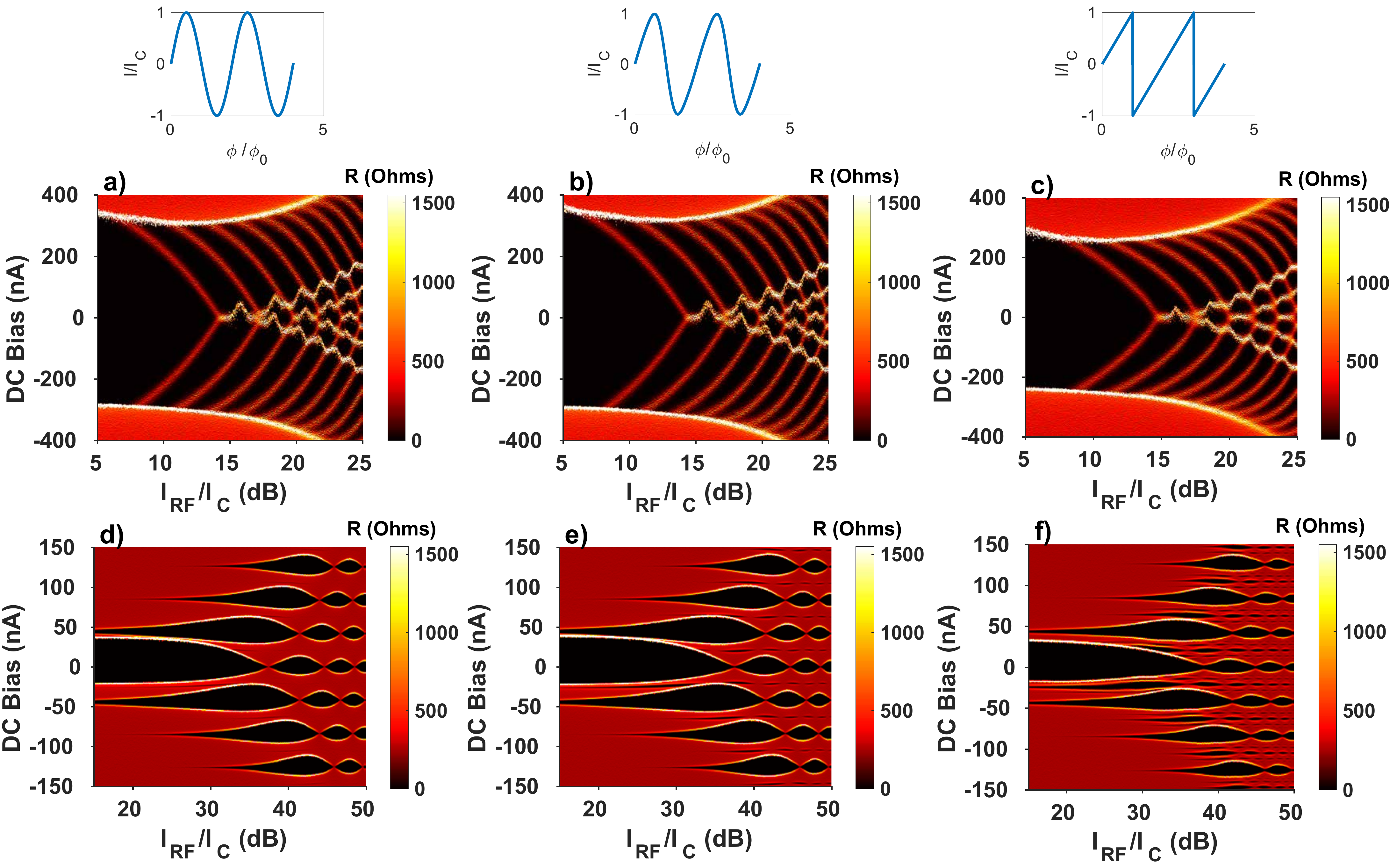}
    \caption{Middle row: Simulations corresponding to Figures 1a and S2a with the skewness of the CPR increasing from left to right (the CPRs are shown in the insets). Even severe skewness does not give rise to fractional steps, although it does slightly alter the map in ways akin to changing parameters such as $I_{c}$ and $R_{j}$. Bottom row: Simulations corresponding to Figures 1d and S2d with the same range of CPR skewness. In this regime, the CPR gives rise to enhanced fractional steps. Top row: CPRs corresponding to the figures below.}
\label{CPRSims}
\end{figure*}

\section{CPR Dependence}
Figure 1d of the main text shows strong fractional Shapiro steps~\cite{AndersonDayem1964}, although there are no signs of fractional steps in the measurements with higher $I_{C}$.  Our simulations are in an agreement with the experimental results, showing that a skewed CPR (current phase relationship) has minimal effect on a sample in the strongly hysteretic regime. Intuitively, we understand the hysteresis of the high $I_{C}$ maps as arising from regions of overlapping stability of integer steps. Thus it may be expected that for such parameters the fractional steps are less stable compared to overlapping integer steps. 
% and therefore not observable on the measurement timescales. \TL{Although simulations suggest they are not stable even on very fast timescales?}

For comparison to the experiment, we took our simulation for Figures 1a and 1d and employed CPRs with varying degrees of skewness~\cite{EnglishCPR2016,Nanda2017}. In the top row, $I_{c}=540$ nA, corresponding to Figures 1a and S2a; in the bottom row, $I_{c}=40$ nA, corresponding to Figures 1d and S2d. The three columns correspond to: sinusoidal CPR (left), a slightly skewed CPR, $I(\phi) = I_{c}[\sin(\phi)-0.2\sin(2 \phi)+0.04\sin(3 \phi)]$ (middle); and a maximally skewed sawtooth CPR (right). The insets in the top panels demonstrate the corresponding CPRs. For large $I_{c}$ simulations, increasing the skewness of the CPR only slightly distorted the map, but did not give rise to any additional plateaus. For small $I_{c}$, increasing the skewness resulted in increasing fractional plateaus. Surprisingly, for small $I_c$ even a perfectly sinusoidal CPR shows some half-quantized steps (Figure S4d). We attribute this behavior to the high frequency environment, which gives rise to some effective skewness. Comparing these simulations to the measured data, we find that the slightly skewed CPR appears to most accurately reproduce the strength of the fractional steps, as expected. 
%In S1b, the RF frequency is still 5 GHz, but the backgate voltage is moved to the Dirac peak.  This gives rise to a small amount of hysteresis is Region I, which is unsurprising given that the quasipotential must be larger as evidenced by the greater amount of hysteresis is region II.

%\end{document}

\end{document}